\documentstyle[prb,aps]{revtex}
\tighten

\begin{document}
\draft
\twocolumn[
\hsize\textwidth\columnwidth\hsize\csname @twocolumnfalse\endcsname

\title{Theory of Andreev reflection in a junction with a strongly
disordered semiconductor}
\author{ I.L. Aleiner, Penny Clarke,  and L.I. Glazman}  
\address{Theoretical Physics Institute, University of Minnesota, 
 Minneapolis, MN 55455}
%\date{Draft: \today}

\maketitle
\begin{abstract}
We study the conduction of a {\sl N~-~Sm~-~S} junction, where {\sl Sm} is a
strongly disordered semiconductor. The differential conductance $dI/dV$ of
this {\sl N~-~Sm~-~S} structure  is predicted to have a sharp peak at
$V=0$.  Unlike the case of a weakly disordered system, this feature
persists even in the absence of an additional (Schottky) barrier on the
boundary. The zero-bias conductance of such a junction $G_{NS}$ is smaller
only by a numerical factor than the conductance in the normal state $G_N$.
Implications for experiments on gated heterostructures with
superconducting leads are discussed.
\end{abstract} 
 ]

\narrowtext
Since the seminal work of Andreev\cite{Andreev} on the theory of electron
transport through an ideal interface between a normal ({\sl N}) metal and a
superconductor ({\sl S}), significant efforts were undertaken to understand
the transport in real {\sl NS} junctions. It was shown\cite{Blonder} that a
barrier at the {\sl N-S} interface reduces strongly the conductance of the
boundary between a clean normal metal and a superconductor. Later,
experiments\cite{Kastalskii} with semiconductor ({\sl Sm}) -- superconductor
junctions revealed in the differential conductance $G_{NS}(V)$ a broad
maximum at zero bias. This feature was explained\cite{Wees0} as an
interference effect due to the scattering off the Schottky barrier and off
the imperfections in the semiconductor. 
Recent technological advances have resulted in fabrication of
low-resistance contacts between a two-dimensional electron gas (2DEG) and a
superconducting lead~\cite{Wees}. Because of the absence of a Schottky
barrier at the interface, the subgap conductance is determined by the
propagation of electron pairs through the 2DEG itself rather than by
two-electron tunneling at the interface, and there is no
peak\cite{Beenakker94} in the differential conductance at zero bias.

The advantage of a gated heterostructure lies in the controllable level of
carrier density in the 2DEG. Depending on the density, the 2DEG may behave
as a good conductor or as an insulator with an adjustable localization
length. Whereas the former case has been extensively studied both
theoretically and experimentally\cite{Beenakker94}, the latter case
has received no attention as of yet. In this paper we study 
two-electron transport through a disordered insulator. We will show that
the zero bias conductance of the  {\sl N~-~Sm~-~S} junction differs from
the conductance of the same structure in the normal state by a numerical
factor only. This is similar to the properties exhibited in the metallic
regime. However, the differential conductance $G_{NS}(V)$ drops abruptly
with increasing voltage, in contrast to the behavior in the metallic
regime in the absence of the Schottky barrier. The development of this
feature of the
$I$-$V$ characteristic under the progressive depletion constitutes the
signature of the crossover between the metallic and insulating regimes.

Deep in the insulating regime, the conductance of the normal 
({\sl N~-~Sm~-~N}) structure
is dominated by tunneling via those configurations of localized states in
the semiconductor layer that facilitate resonant transmission of
electrons\cite{Beasley}. An example of such a configuration is a
state with  energy close to the Fermi level and location symmetric
with respect to the leads. The transmission coefficient for an electron
tunneling through a resonant configuration is close to unity, and
the conductance is proportional to the probability $w$ of finding such a
configuration.This probability scales exponentially with the length of the
{\sl Sm} region, $L$. 

The zero-bias conductance of the  {\sl N~-~Sm~-~S} junction, $G_{NS}(0)$,
is determined by the tunneling of pairs of electrons at the Fermi
level\cite{Andreev}. Clearly, these tunneling processes are also
facilitated by the same resonant configurations that control 
single-electron transport. Thus, the conductance $G_{NS}(0)$ is also
proportional to $w$ and, therefore, it has the same exponential dependence
upon $L$ as the does normal conductance. If a finite bias $eV$ is applied
to the junction, the energies $\epsilon_1$, $\epsilon_2$ of the two 
electrons in the pair are different:  $\epsilon_1 -
\epsilon_2 =  2 eV$. If
$eV$ exceeds the width of the resonant level with respect to tunneling,
this level cannot provide a large tunneling coefficient for both electrons.
It results in a sharp drop of the conductance with voltage. 

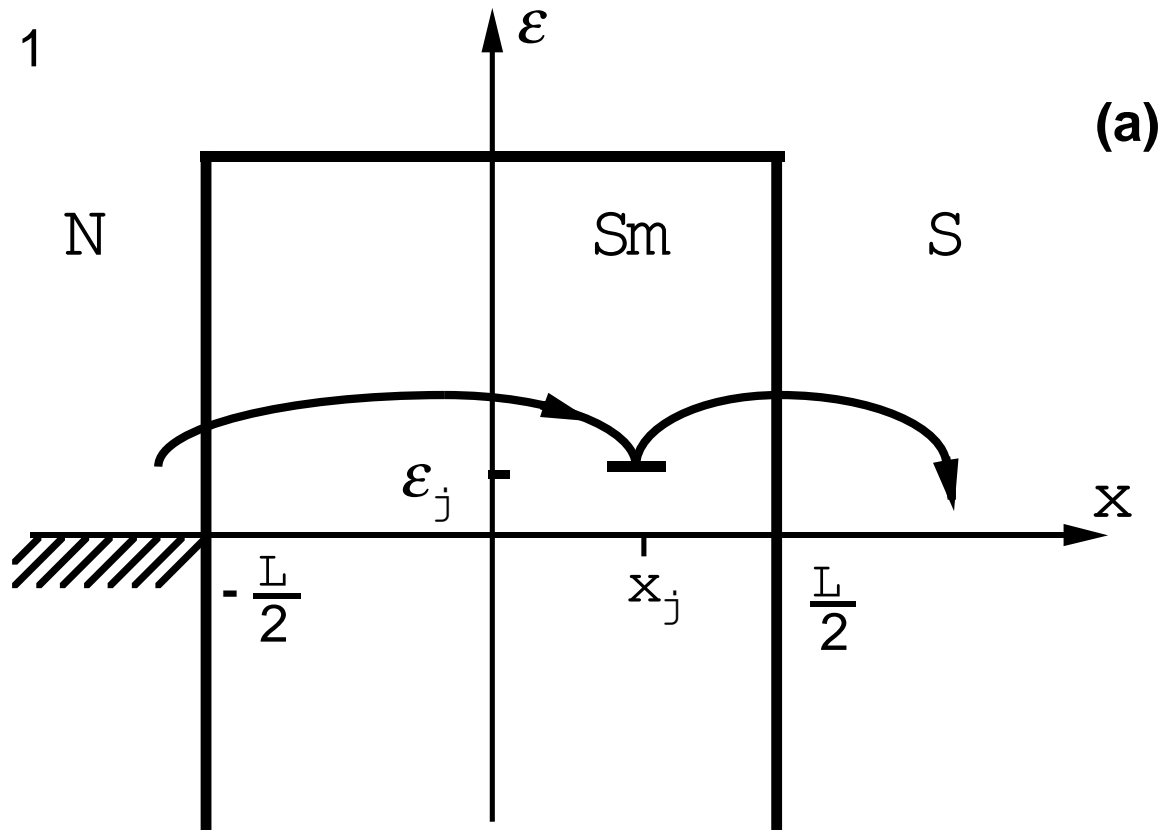
\begin{figure}
\caption{Schematic picture of the {\sl N~-~Sm~-~S} junction with (a)
single localized state, (b) two-impurity chains}
\label{fig:1}
\end{figure}

Following Ref.\onlinecite{Matveev88}, we model transport through
the depleted region as  resonant tunneling via isolated localized states
(impurities). We will show that for a wide range of lengths $L$, it
suffices to consider  the single impurity configurations only. In order to
calculate the conductance, we first calculate the contribution to the
conductance due to  tunneling through a single impurity
$g_{NS}(eV)$ and then sum these partial conductances over all the
impurities. Each localized state is characterized by its energy
$\epsilon_j$ and by the level widths  $\Gamma_{l(r)}^{(j)}$ due to the
decay into the  left (right) lead, see Fig.~\ref{fig:1}a. The amplitude of
 electron transmission through the barrier via the resonant state,
$t_{(j)}(\epsilon)$, and the amplitude of reflection, $r_{(j)}(\epsilon)$,
at energy  $\epsilon$ are given by the single channel Breit-Wigner formula:
\begin{eqnarray}
r_{(j)}(\epsilon )=\!  
\frac{\epsilon - \epsilon_j + i \left(\Gamma_l^{(j)} -
\Gamma_r^{(j)}\right)} {\epsilon - \epsilon_j + i \left(\Gamma_l^{(j)} +
\Gamma_r^{(j)}\right)},
\nonumber \\  t_{(j)}(\epsilon )=\!\frac{-2 i
\sqrt{\Gamma_l^{(j)}\Gamma_r^{(j)}}} {\epsilon - \epsilon_j + i
\left(\Gamma_l^{(j)} + \Gamma_r^{(j)}\right)}.
\label{oneimpurity}
\end{eqnarray} The tunneling widths $\Gamma_{l(r)}^{(j)}$ depend
exponentially on the distance of the impurity from the middle of the
barrier, $x_j$, (see also Fig.~\ref{fig:1}a):
\begin{equation}
\Gamma_l^{(j)} = E_0 e^{-(L+2x_j)/{a_0}},\quad
\Gamma_r^{(j)} = E_0 e^{-(L-2x_j)/{a_0}}, 
\label{rates}
\end{equation}
where $a_0$ is the localization radius of the impurity state, and the
energy $E_0$ can be estimated as
$E_0 \simeq \hbar^2/ma_0^2$, with $m$ being the electron mass in the Sm
layer. The Andreev reflection probability can be
expressed\cite{Beenakker94} in terms of the one-electron amplitudes
$t_{(j)}(\epsilon)$ and $r_{(j)}(\epsilon)$. The corresponding
contribution $g_{NS}(eV;x_j,\epsilon_j)$ of  localized state $j$ to the
conductance is:
\begin{equation}
g_{NS}(eV;x_j,\epsilon_j) =
\frac{2e^2}{\pi\hbar}\left|\frac{t_{(j)}(eV)t_{(j)}^\ast (-eV)}{1 +
r_{(j)}(eV)r_{(j)}^\ast (-eV)}\right|^2.
\label{gnsdefinition}
\end{equation}
(We will restrict our discussion to the most interesting regime of the
bias being small compared to the superconducting gap.)

Now we  sum up the contributions to the conductance from different
impurities. Assuming that the density of the localized states $\rho$ is
independent of energy, we obtain:
\begin{equation}
G_{NS}^{(1)}(eV) = \rho W
\int_{-\infty}^{\infty}\!d\epsilon_j\int_{-L/2}^{L/2}\!\!dx_j 
\ g_{NS}(eV;x_j,\epsilon_j).
\label{NSaverage}
\end{equation}
Here $W$ is the width of the barrier. 

The calculation of the conductance is thus reduced to the 
integration in Eq.~(\ref{NSaverage}), with the help of formulas
(\ref{oneimpurity}) -- (\ref{gnsdefinition}). In the low bias limit, $eV
\ll
\Gamma_1$, we find
\begin{equation}
\!\!G_{NS}^{(1)}\! = \!\frac{e^2}{\hbar}\!\left(\rho a_0W\Gamma_1\right) 
\left[\frac{\Gamma
(3/4)^2}{\sqrt{\pi}}+ \frac{\Gamma
(1/4)^2}{96\sqrt{\pi}} \left(\frac{eV}{\Gamma_1}\right)^2\right],
\label{lowbias}
\end{equation}
where $\Gamma_1\equiv E_0e^{-L/a_0}$ is the level width of a state
localized at $x_j=0$, and $\Gamma (x)$ is the Gamma
function.  It is instructive to express this result in terms of the normal
state conductance of the same junction:
\begin{eqnarray}
&\displaystyle{G_{N}^{(1)} = \frac{e^2\rho W}{\pi\hbar}
\int_{-\infty}^{\infty}d\epsilon_j\int_{-L/2}^{L/2}dx_jg_{N}(eV;x_j,
\epsilon_j)}&,\nonumber\\
&\displaystyle{g_{N}(eV;x_j,\epsilon_j)=\frac{e^2}{\pi\hbar}|t_{(j)}|^2.}&
\label{Naverage}
\end{eqnarray}
A simple calculation based on (\ref{oneimpurity}) and
(\ref{Naverage})  gives\cite{Matveev88,2D}
\begin{equation}
G_{N}^{(1)} = \frac{e^2}{\hbar} \left(\pi\rho a_0W\Gamma_1\right).
\label{Nresult}
\end{equation}
Comparing Eqs.~(\ref{lowbias}) and (\ref{Nresult}) we obtain:
\begin{equation}
G_{NS}^{(1)} = G_{N}^{(1)}\left[0.27 + 0.049 
\left(\frac{eV}{\Gamma_1}\right)^2\right].
\label{lowbiasnumbers}
\end{equation}

Results (\ref{lowbias}) and (\ref{Nresult}) can be easily understood. The
contribution of the single-site resonant states to both $G_N$ and
$G_{NS}$ is determined by the number of the states with the energy 
near the Fermi level within the strip of width $\Gamma_1$ and positioned
within the strip of the width $a_0$ around the middle of the barrier, so
that $\Gamma_l \simeq \Gamma_r$. Therefore, the factor $\rho a_0W\Gamma_1$
is just the number of such states. Result (\ref{lowbiasnumbers}) at $V=0$
was obtained independently in Ref.~\onlinecite{Khlus}.

Let us estimate the domain of parameters within which  the mehanism of
tunneling through rare single resonant states dominates over  direct
tunneling through the potential barrier created by the depleted {\sl Sm}
layer. The contribution of the latter mechanism can be estimated as
$G_N^{dir}
\sim (e^2/\hbar) (k_F W) e^{-2L/a_0}$, where $k_F$ is the Fermi wave
vector in the {\sl N} lead. Thus direct tunneling is irrelevant for not too
short barriers: 
\begin{equation}
L \gg a_0 \ln \left(\frac{ma_0k_F}{\hbar^2\rho} \right).
\label{impdirect}
\end{equation}
This condition is easily met in the experimental situation
\cite{Beasley,Washburn}.

The linear conductance, $G_{NS}^{(1)}(0)$, differs only by a numerical
factor from $G_{N}^{(1)}$ because the optimal configurations contributing
to both quantitites are the same. It changes drastically when the the 
bias increases, $eV\gg\Gamma_1$. In this regime we obtain from
Eq.~(\ref{NSaverage})
\begin{equation}
G_{NS}^{(1)}(V) = G_{N}^{(1)}\frac{\Gamma_1}{eV}.
\label{highbias}
\end{equation} 
The main contribution to $G_{NS}^{(1)}(V)$ comes not from the impurities
located  near the middle of the barrier but rather from ones shifted closer
to the {\sl S} lead.

Result (\ref{highbias}) can be understood using the following arguments.
Consider the impurity lying at the Fermi level, $\epsilon_j=0$, that is
completely  decoupled from the normal lead, $\Gamma_l=0$.  The tunneling of
an electron pair between the superconductor and the impurity mixes the
states
$|0\rangle$ and $|2\rangle$, corresponding to zero and two electrons
occupying the impurity.
As the result, the four-fold degeneracy of the impurity level is
partially lifted: two singly occupied states still have energy
$\epsilon_j=0$, but the other two states, $\left|0\right>-\left|2\right>$
and $\left|0\right>+\left|2\right>$, are split by $2\Gamma_r$
symmetrically with respect to the Fermi level. The even wave function is
the ground state of the impurity, and the smallest excitation energy
(to the singly occupied state) is $\Gamma_r$. Let us now turn on
a small coupling between the impurity and the normal lead, $\Gamma_l \ll
\Gamma_r$. This coupling enables the electron to tunnel from the Fermi
level in {\sl N} lead to the impurity level. This tunneling is
resonant if the potential drop on the contact
$eV$ is close to the energy of excitation, $|eV -\Gamma_r| \lesssim
\Gamma_l$. A similar consideration is valid for all the impurities with
energies
$|\epsilon_j|\lesssim \Gamma_r$.  Now, we find the impurities that
give the maximal contribution to the conductance. The transmission
coefficient is close to unity for the impurities with $\Gamma_r = eV$,
i.e. at $eV>\Gamma_1$ the optimal impurities are shifted towards the
superconducting lead. The energies of these impurities may lie in the strip
of the width
$\Gamma_r$ about the Fermi level. The coordinate $x_j$ of the impurity may
deviate from its optimal value by no more than
$\pm a_0\Gamma_l/\Gamma_r$, as it follows from the condition 
$|eV -\Gamma_r| \lesssim \Gamma_l$, and Eq.~(\ref{rates}). This
consideration shows that the number of  relevant impurities is of the
order of $\rho a_0 \Gamma_l \Gamma_r/eV$, which with the help of
Eqs.~(\ref{rates}) and (\ref{Nresult}) immediately yields
Eq.~(\ref{highbias}).

As the barrier length is increased, the dominant mechanism of
electron transport shifts from the single-state configurations to the
configurations containing ``chains'' of two localized
states\cite{Matveev88}.  These configurations are more complex than the
ones considered before, nevertheless Eq.~(\ref{gnsdefinition}) is still
valid. The only difference is that the one electron reflection and
transmission amplitudes now depend upon the energies
$\epsilon_{j_l}, \epsilon_{j_r}$ and positions
$\mbox{\boldmath $r$}_{j_l},\mbox{\boldmath $r$}_{j_r} $ of the two
impurities, see Fig.~\ref{fig:1}b. Each impurity is characterized by its
coupling to the {\em nearest} lead, see  Fig.~\ref{fig:1}b,
\begin{equation}
\Gamma^{(j_l)} = E_0 e^{-(L+2x_{j_l})/{a_0}},\quad
\Gamma^{(j_r)} = E_0 e^{-(L-2x_{j_r})/{a_0}}. 
\label{rates2}
\end{equation}
The coupling between the two impurities is given by
\begin{equation}
h_{j_lj_r} = {E_1}\left(\frac{a_0}{\left|\mbox{\boldmath $r$}_{j_l}-
\mbox{\boldmath $r$}_{j_r}\right|}\right)^{1/2}e^{-\left|
\mbox{\boldmath $r$}_{j_l}-
\mbox{\boldmath $r$}_{j_r}\right|/a_0},
\label{h12}
\end{equation}
where the energy $E_1$ is of the order of $E_0$.

After one finds the partial con\-duc\-tance
${g}_{NS}$ of a single two-impurity ``chain'', the net
contribution ${G}_{NS}^{(2)}(eV)$ of these chains to the total conductance
can be calculated in a  manner similar  to that used previously for the
single-impurity configurations:
\begin{eqnarray}
&&\displaystyle{G_{NS}^{(2)}(eV) = \rho^2W\!
\int\!\!\!\!\int\!\!\!\!\int_{-\infty}^{\infty}\!\!d\epsilon_{j_r}
d\epsilon_{j_l}dy_{j_l}
\!\int_{-L/2}^{L/2}\!\!dx_{j_r}\!\int^{x_{j_r}}_{-L/2}\!\!dx_{j_l}}
\nonumber\\
&&\displaystyle{\hspace{1cm}\times {g}_{NS}(eV;
\epsilon_{j_r},\epsilon_{j_l},x_{j_l},x_{j_r}, y_{j_l}-y_{j_r}).}
\label{NS2average}
\end{eqnarray}

The formulas for the reflection and transmission amplitudes entering 
${g}_{NS}$ for an arbitrary impurity pair are quite cumbersome.
Fortunately, the dominant part of the
average (\ref{NS2average}) comes from the impurity pairs with
sufficiently large energy difference,
$|\epsilon_{j_l} - \epsilon_{j_r}| \gg h_{j_lj_r}$. This means that one of
the two impurities serves as a resonant level for the incoming
electron, while the second impurity provides a virtual state which
modifies the escape rate from the resonant level into the lead.
Therefore, for the relevant ``chains'' the reflection and transmission
amplitudes are given by Eq.~(\ref{oneimpurity}) with the level widths
renormalized by the tunneling through a virtual state. Depending on which
component of the pair is in resonance, we find:
\begin{mathletters}
\begin{equation}
\epsilon_{j} \to \epsilon_{j_l},\  \Gamma^{(j)}_l \to \Gamma^{(j_l)},
\  \Gamma^{(j)}_r
\to \Gamma^{(j_r)}\left(\frac{h_{j_lj_r}}
{\epsilon_{j_l}-\epsilon_{j_r}}\right)^2
\label{substitution1}
\end{equation}
if the left impurity is in resonance, or
\begin{equation}
\epsilon_{j} \to \epsilon_{j_r},\  \Gamma^{(j)}_l \to
\Gamma^{(j_l)}\left(\frac{h_{j_lj_r}}
{\epsilon_{j_l}-\epsilon_{j_r}}\right)^2,
\  \Gamma^{(j)}_r
\to \Gamma^{(j_r)}
\label{substitution2}
\end{equation}
\label{substitution}
\end{mathletters}
\noindent
if the right impurity is the resonant one. The corresponding contributions
to ${g}_{NS}$ are obtained from Eqs.~(\ref{gnsdefinition})
and (\ref{oneimpurity}) by the substitutions (\ref{substitution1}) and
(\ref{substitution2}) respectively. 

Similar to the case of single impurity configurations, it is convenient
to express the result in terms of the contribution of two-impurity chains
to the normal  conductance. This contribution
${G}_{N}^{(2)}$, is given by\cite{Matveev88}
\begin{eqnarray}
&&\displaystyle{{G}_{N}^{(2)}(eV) = \rho^2W\!
\int\!\!\!\!\int\!\!\!\!\int_{-\infty}^{\infty}\!\!d\epsilon_{j_r}
d\epsilon_{j_l}dy_{j_l}
\!\int_{-L/2}^{L/2}\!\!dx_{j_r}\!\int^{x_{j_r}}_{-L/2}\!\!dx_{j_l}}
\nonumber\\
&&\displaystyle{\hspace{1cm}\times {g}_{N}(eV;
\epsilon_{j_r},\epsilon_{j_l},x_{j_l},x_{j_r}, y_{j_l}-y_{j_r}),}
\label{N2average}
\end{eqnarray}
where the partial conductances ${g}_{N}$ are 
obtained from Eqs.~(\ref{Naverage}) and (\ref{oneimpurity}) by the 
substitution of (\ref{substitution1}) and (\ref{substitution2}).
Simple integration in Eq.~(\ref{N2average}) yields
\begin{equation}
{G}_N^{(2)} = 3\sqrt{\frac{\pi}{2}}\rho L^2E_1G_N^{(1)},
\label{N2result}
\end{equation}
where $G_N^{(1)}$ is the single impurity result Eq.~(\ref{Nresult}).
Two-impurity chains dominate the conductance\cite{Matveev88,2D} if the
disordered region is long enough:
\begin{equation}
L \gtrsim a_0\left(\frac{m}{\hbar^2\rho} \right)^{1/2}.
\label{double}
\end{equation}
(We used here the estimate $E_1\sim E_0\sim\hbar^2/ma_0^2$.) Deep in the
insulating regime, where the separate impurity states overlap weakly, the 
parameter $\hbar^2\rho/m$ is small, and the condition (\ref{double}) is
more restrictive than (\ref{impdirect}).

If the length of the barrier increases further, more complex 
configurations may contribute. The effects associated with these
configurations have not been observed experimentally for the
normal conductance and  we will not consider their contributiton to
$G_{NS}$.

Now we present our results for the two-impurity contribution to 
{\sl N~-~Sm~-~S} conductance obtained by performing the integration in
Eq.~(\ref{NS2average}).  At zero-bias, we found 
\begin{equation}
G_{NS}^{(2)}(0) = \frac{\Gamma
(3/4)^2}{\sqrt{\pi}}G_{N}^{(2)} \approx 0.27 G_{N}^{(2)}.
\label{lowbias2}
\end{equation}
It is important to emphasize that the relationships between the
contributions to the {\sl N~-~Sm~-~S} and {\sl N~-~Sm~-~N} conductances are
identical for one- and two-impurity configurations, see
Eq.~(\ref{lowbiasnumbers}). This is because within the optimal two-impurity
chains only one level is in resonance, and the other level is
responsible only for the tunneling width.  The same argument also
apply for the one dimensional chains consisting of a larger number of
impurities. Therefore, we believe, that the relation ${G}_{NS}(0)
=0.27{G}_{N}$ is a universal property of disordered junctions for which the
conductivity occurs via the tunneling through quasi-one-dimensional
chains containing an arbitrary number of impurities. 

For a finite bias, however, the relationship beween $G_{NS}^{(2)}$
and $G_N^{(2)}$ is quite different from Eq.~(\ref{highbias}): the
differential conductance $G_N^{(2)}$ drops only logarithmically with the
increase of $V$. The strongest variation of the conductance,
\begin{equation}
G_{NS}^{(2)}(eV) = G_{NS}^{(2)}(0)\frac{4
a_0^2}{3L^2}\ln^2\left(\frac{\Gamma_2}{eV}\right),
\label{highbias2}
\end{equation}
occurs in the region $\Gamma_1\lesssim eV\lesssim\Gamma_2$. Here
$\Gamma_2\equiv E_0\exp(-L/2a_0)=\Gamma_1\exp(L/2a_0)$ is the
characteristic width of a resonant level formed by a two-impurity chain.

The reason for the logarithmic dependence (\ref{highbias2}) is the
following. As we already discussed, the transmission coefficient for 
two electron tunneling through the localized level is close to unity, if
two conditions are met: (i) the tunneling rates $\Gamma_l, \Gamma_r$ are
approximately the same, and (ii) the difference between the energies of
the two tunneling electrons is  less than the level width, $eV <
\Gamma_{l,r}$. For the single impurity configuration, the first condition
can be met only for the impurities located in the vicinity of the middle
of the barrier, therefore, the sharp drop occurs at $eV \sim \Gamma_1$.
However, for the two impurity configuration, the first condition can be
satisfied for various configurations $\{x_{j_l}, x_{j_r}\}$. The largest
level width compatible with the condition (i), $\Gamma_2=\Gamma_1
e^{L/2a_0}$, corresponds to the largest separation $X=L/2$ between the two
impurities. On the other hand, condition (ii) restricts the separation
from below,  $X\gtrsim a_0\ln(eV/\Gamma_1)$, which eventually leads to
Eq.~(\ref{highbias2}).
 
Comparison of Eq.~(\ref{highbias}) with Eq.~(\ref{highbias2}) shows, that
even if at low biases conductance is dominated by single-impurity channels,
a crossover to the two impurities``chain'' configurations may occur at
larger
$eV$. This crossover  from a sharp function (\ref{highbias}) to a much
slower logarithmic dependence (\ref{highbias2}) takes place at voltage
\[
eV^\ast\simeq\frac{e^{-\lambda}}{\lambda^2}\Gamma_2,
\quad\lambda \equiv \frac{L}{2a_0}-\ln\left(\frac{m}{\hbar^2\rho} \right).
\]
At this bias, the optimal configurations changes from single impurity to
two impurities ``chains''.
To observe the crossover, the junction parameters should satisfy the
condition:
\[
1\ll\lambda\ll\left(\frac{m}{\hbar^2\rho} \right)^{1/2}.
\]

In conclusion, we studied the conductance of the {\sl N~-~Sm~-~S}
junction where {\sl Sm} is a strongly disordered semiconductor. The
electron transport is due to resonant tunneling through the levels
localized in {\sl Sm}. We find, that at zero-bias {\sl N~-~Sm~-~S}
conductance is proportional to the conductance of the same junction in the
normal state,
$G_{NS}(0) = 0.27 G_N$. At larger biases, the conductance $G_{NS}(V)$ drops
drastically. This drop represents the signature of Andreev reflection
in a junction with a strongly disordered semiconductor. 
  
Discussions with F.W.J. Hekking are acknowledged with gratitude.
This work was partially supported by the Minnesota Supercomputer
Institute and by NSF Grant DMR-9423244.

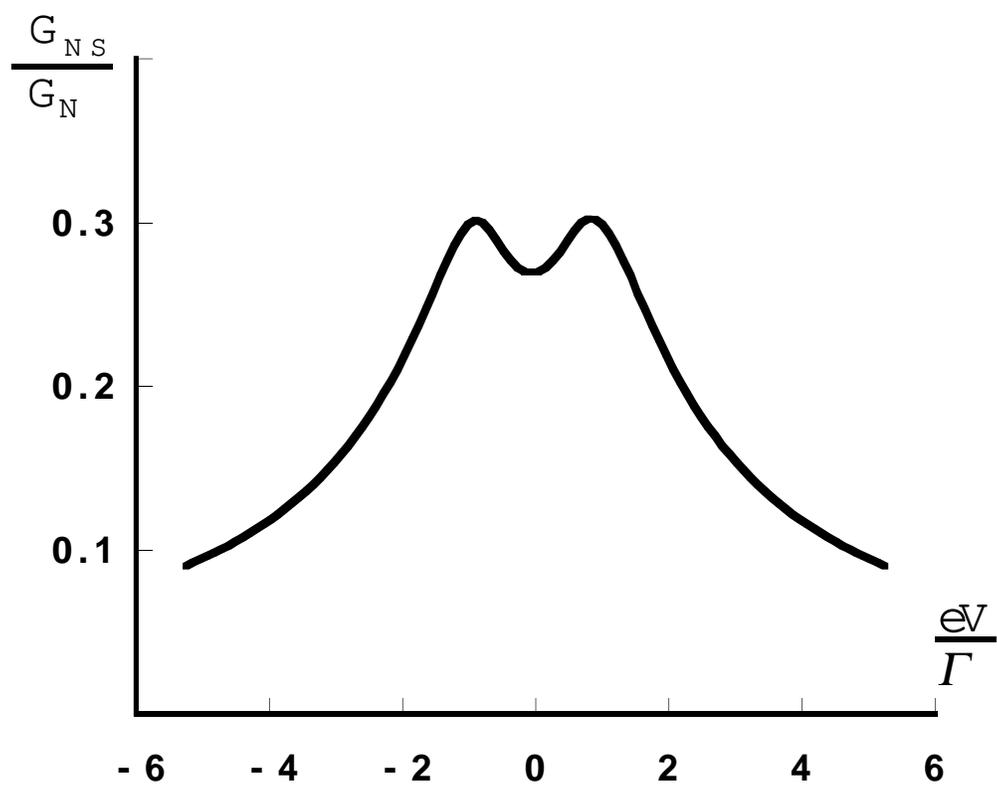
\begin{figure}
\caption{Voltage dependence of the differential {\sl N~-~Sm~-~S}
conductance  contributed by single impurity configurations,
$G_{NS}^{(1)}(V)$.}
\label{fig:2}
\end{figure}


\begin{references}
\bibitem{2D} 
 The difference between our expressions for
the normal conductance and those from Ref.\onlinecite{Matveev88} is due
to the 2D depleted layer in our case, whereas 3D amorphous barier was
considered in Ref.\onlinecite{Matveev88}.
\end{references}
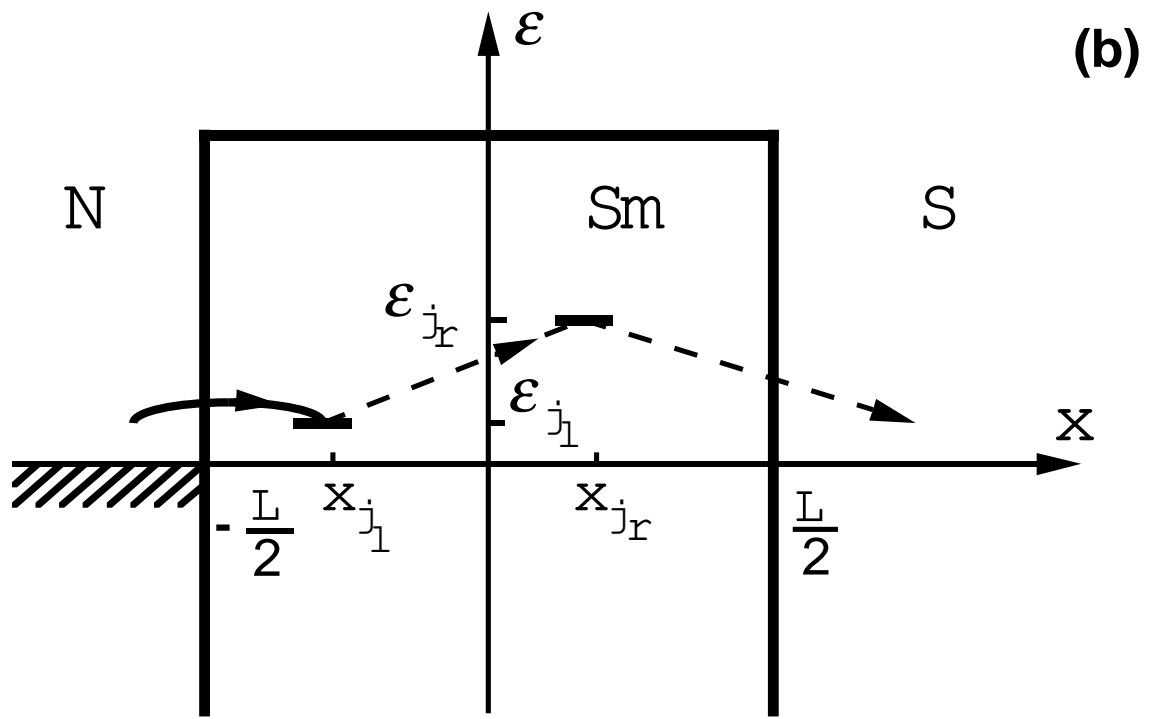
\end{document}